\documentclass[pra,aps,twocolumn,floatfix,showpacs]{revtex4}
\usepackage{graphicx,amsmath,amssymb,times}
\begin{document}
\title{Rotation induced superfluid-normal phase separation in trapped Fermi gases}
\author{M. Iskin and E. Tiesinga}
\affiliation{Joint Quantum Institute, National Institute of Standards and Technology, and 
University of Maryland, Gaithersburg, Maryland 20899-8423, USA.}
\date{\today}

\begin{abstract}

We use the Bogoliubov-de Gennes formalism to analyze the effects of
rotation on the ground state phases of harmonically trapped Fermi gases,
under the assumption that quantized vortices are not excited. 
We find that the rotation breaks Cooper pairs that are located near the trap edge, 
and that this leads to a phase separation between the nonrotating superfluid 
(fully paired) atoms located around the trap center and the rigidly rotating 
normal (nonpaired) atoms located towards the trap edge, with a coexistence 
(partially paired) region in between. Furthermore, we show that the 
superfluid phase that occurs in the coexistence region is characterized by 
a gapless excitation spectrum, and that it is distinct from the gapped phase 
that occurs near the trap center.

\end{abstract}

\maketitle

With the ultimate success of techniques for trapping and cooling atomic 
gases developed and improved gradually since the 1980s, atomic Fermi 
gases have emerged as unique testing grounds for many theories of 
exotic matter in nature, allowing for the creation of complex yet very 
accessible and controllable many-body quantum systems. For instance, 
evolution from the weakly attracting Bardeen-Cooper-Schrieffer (BCS) 
limit to the weakly repulsive molecular Bose-Einstein condensation (BEC) 
have been observed in a series of remarkable experiments. The ground
state phases of such atomic Fermi gases have since been the subject of 
intense theoretical and experimental research worldwide~\cite{varenna, giorgini}.

To verify the superfluid ground state of atomic Fermi gases, it is essential 
to analyze their response to rotation. 
For instance, it has been theoretically predicted that sufficiently fast rotation
of superfluid Fermi gases excites quantized vortices in the form of hexagonal
vortex lattices~\cite{nygaard, machida, sensarma}. Such vortex lattices have 
recently been observed for both population balanced~\cite{mit-vortex} 
and imbalanced~\cite{mit-mixture} systems. These vortex experiments 
have not only complemented previously found signatures, but also 
provided the ultimate evidence to support the superfluid nature of the ground state.

Recently, the effects of rotation on the ground state phases of 
harmonically trapped Fermi gases has been theoretically studied at unitarity, 
under the assumption that quantized vortices are not excited~\cite{stringari}
(see also Ref.~\cite{hui}).
It has been argued that the rotation causes a complete phase separation 
between a nonrotating superfluid core and a rigidly rotating normal gas,
with a discontinuous density at the interface. Subsequently, this problem has 
been analyzed using the mean-field BCS framework 
within the semi-classical local density approximation (LDA)~\cite{urban}. 
In addition to the superfluid and normal regions, a coexistence region is
found, the possibility of which is not considered in Ref.~\cite{stringari}.

In this manuscrript, we go beyond the semi-classical LDA and develop a quantum 
mechanical Bogoliubov-de Gennes (BdG) formalism to analyze the 
effects of rotation on the ground state phases of harmonically 
trapped Fermi gases. We discuss both population balanced and imbalanced
mixtures throughout the BCS-BEC evolution.
Our main results are in qualitative agreement with those of Ref.~\cite{urban}.
We find that the rotation breaks Cooper pairs that are located near the trap edge, 
and that this leads to a phase separation between the nonrotating superfluid 
(fully paired) atoms located around the trap center and the rigidly rotating normal 
(nonpaired) atoms located towards the trap edge, with a coexisting 
(partially paired) region in between. This leads to a continuous density and 
superfluid order parameter as a function of radial distance.
Furthermore, we show that the superfluid phase that 
occurs in the coexistence region is characterized by a gapless
excitation spectrum, and that it is distinct from the gapped phase that occurs 
near the trap center.

We obtain these results by solving the mean-field BdG equations in the rotating 
frame (in units of $\hbar = k_B = 1$)
\begin{equation}
\label{eqn:bdg}
\left[ \begin{array}{cc}
{\cal K}_\uparrow(\mathbf{r}) - \Omega {\cal L}_z & \Delta(\mathbf{r}) \\
\Delta^*(\mathbf{r}) & -{\cal K}_\downarrow^*(\mathbf{r}) + \Omega {\cal L}_z^*
\end{array} \right]
\left[ \begin{array}{c} u_\eta(\mathbf{r}) \\ v_\eta(\mathbf{r}) \end{array} \right]
= \epsilon_{\eta} \left[ \begin{array}{c} u_\eta(\mathbf{r}) \\ v_\eta(\mathbf{r}) \end{array} \right],
\end{equation}
where 
$
{\cal K}_\sigma (\mathbf{r}) = -\nabla^2/(2M) - \mu_\sigma(r),
$
$\Omega$ is the rotation frequency around the $z$-axis of the trapping potential, 
${\cal L}_z$ is the $z$-component of the angular momentum operator, 
and the off-diagonal self-consistency field
$
\Delta(\mathbf{r}) = g \langle \psi_\uparrow(\mathbf{r}) \psi_\downarrow(\mathbf{r}) \rangle
$
is the local superfluid order parameter. 
Here, $\sigma \equiv \{ \uparrow, \downarrow\}$ labels the trapped hyperfine states,
$M$ is the mass, $\mu_\sigma(r) = \mu_\sigma - V(r)$ 
is the local chemical potential, $\mu_\sigma$ is the global chemical potential, 
$V(r) = M \omega^2 r^2/2$ is the trapping potential which is assumed to be spherically 
symmetric, $\omega$ is the trapping frequency, $g > 0$ is the strength of the 
zero-ranged attractive interactions between $\uparrow$ and $\downarrow$ atoms,
and $\langle ... \rangle$ is a thermal average.
The quasiparticle wavefunctions $u_{\eta}(\mathbf{r})$ and $v_{\eta}(\mathbf{r})$ are
related to the particle annihilation operator $\psi_\sigma(\mathbf{r})$ via 
the Bogoliubov-Valatin transformation
$
\psi_\sigma(\mathbf{r}) = \sum_\eta[u_{\eta,\sigma}(\mathbf{r}) \gamma_{\eta,\sigma} 
- s_\sigma v_{\eta,\sigma}^*(\mathbf{r}) \gamma_{\eta,-\sigma}^\dagger],
$
where $\gamma_{\eta,\sigma}^\dagger$ and $\gamma_{\eta,\sigma}$ are the
quasiparticle creation and annihilation operators, respectively, 
and $s_\uparrow = +1$ and $s_\downarrow = -1$.
Since the BdG equations are invariant under the 
transformation $v_{\eta,\uparrow}(\mathbf{r}) \to u_{\eta,\uparrow}^*(\mathbf{r})$,
$u_{\eta,\downarrow}(\mathbf{r}) \to -v_{\eta,\downarrow}^*(\mathbf{r})$ and 
$\epsilon_{\eta,\downarrow} \to -\epsilon_{\eta,\uparrow}$, it is sufficient to solve only 
for $u_\eta(\mathbf{r}) \equiv u_{\eta,\uparrow}(\mathbf{r})$, 
$v_\eta(\mathbf{r}) \equiv v_{\eta,\downarrow}(\mathbf{r})$ and 
$\epsilon_\eta \equiv \epsilon_{\eta,\uparrow}$ as long as we keep all of the 
solutions with positive and negative eigenvalues.

We assume
$
\Delta(\mathbf{r}) = - g \sum_{\eta} u_{\eta}(\mathbf{r}) v_{\eta}^*(\mathbf{r}) f(\epsilon_{\eta})
$
is real without loosing generality, where $f(x) = 1/[\exp(x/T) + 1]$ is the Fermi 
function and $T$ is the temperature. 
Furthermore, we can relate $g$ to the two-body scattering length $a_F$ via
$
1/g = -M/(4\pi a_F) + M k_c(r)/(2\pi^2) 
$
where
$
k_c^2(r)= 2M[\epsilon_c + \mu(r)]
$
and $\mu(r) = [\mu_\uparrow(r) + \mu_\downarrow(r)]/2$.
Here, $\epsilon_c$ is the energy cutoff to be specified below, and our results depend 
weakly on the particular value of $\epsilon_c$ provided that it is chosen sufficiently high. 
The order parameter equation has to be solved self-consistently with the number equation
$
N_\sigma = \int d\mathbf{r} n_\sigma(\mathbf{r}),
$
where
$
n_\sigma(\mathbf{r}) = \langle \psi_\sigma^\dagger(\mathbf{r}) \psi_\sigma(\mathbf{r}) \rangle
$
is the local density of fermions, leading to
$
n_\uparrow(\mathbf{r}) = \sum_{\eta} |u_\eta(\mathbf{r})|^2 f(\epsilon_\eta)
$
and
$
n_\downarrow(\mathbf{r}) = \sum_{\eta} |v_\eta(\mathbf{r})|^2 f(-\epsilon_\eta).
$

Next, we expand $u_\eta(\mathbf{r})$ and $v_\eta(\mathbf{r})$ in the complete basis 
of the harmonic trapping potential eigenfunctions given by
$
{\cal K}_\sigma(\mathbf{r}) \phi_{n,\ell,m} (\mathbf{r}) 
= \xi_{n,\ell}^\sigma \phi_{n,\ell,m} (\mathbf{r}),
$
where 
$
\xi_{n,\ell}^\sigma = \omega(2n+\ell+3/2) - \mu_\sigma
$
is the eigenvalue and
$
\phi_{n,\ell,m} (\mathbf{r}) = R_{n,\ell} (r) Y_{\ell,m}(\theta_\mathbf{r},\varphi_\mathbf{r})
$
is the eigenfunction. Here, $n$ is the radial quantum number, and $\ell$ and $m$
are the orbital angular momentum and its projection, respectively.
The angular part $Y_{\ell,m}(\theta_\mathbf{r},\varphi_\mathbf{r})$ 
is a spherical harmonic and the radial part is
$
R_{n,\ell} (r) = \sqrt{2} (M \omega)^{3/4} [n!/(n+\ell+1/2)!]^{1/2}
e^{-\bar{r}^2/2} \bar{r}^\ell L_n^{\ell+1/2}(\bar{r}^2),
$
where 
$
\bar{r} = \sqrt{M \omega} r
$ 
is dimensionless and $L_i^j(x)$ is an associated Laguerre 
polynomial. 

We choose ${\cal L}_z = -i \partial/\partial \varphi_{\mathbf{r}}$ and 
$\eta \equiv \{\ell, m, \gamma\}$, leading to
$
u_{\ell,m,\gamma} (\mathbf{r}) = \sum_n c_{\ell,m,\gamma,n} \phi_{n,\ell,m} (\mathbf{r})
$
and
$
v_{\ell,m,\gamma} (\mathbf{r}) = \sum_n d_{\ell,m,\gamma,n} \phi_{n,\ell,m} (\mathbf{r}).
$
This expansion reduces the BdG equations to a $2(n_\ell+1) \times 2(n_\ell+1)$ 
matrix eigenvalue problem for a given $\{\ell, m\}$ state
\begin{align}
\label{eqn:bdg.matrix}
\sum_{n'} \left( \begin{array}{cc}
K_{\uparrow, \ell}^{n,n'} -  m \Omega \delta_{n,n'} & \Delta_\ell^{n,n'} \\
\Delta_\ell^{n',n} & -K_{\downarrow, \ell}^{n,n'} - m \Omega \delta_{n,n'} 
\end{array} \right)
& \left( \begin{array}{c}
c_{\ell,m,\gamma,n'} \\
d_{\ell,m,\gamma,n'} 
\end{array} \right)  \nonumber \\
=\epsilon_{\ell,m,\gamma}
\left( \begin{array}{c}
c_{\ell,m,\gamma,n} \\
d_{\ell,m,\gamma,n} 
\end{array} \right).
&
\end{align}
Here, $n_\ell = (n_c - \ell)/2$ is the maximal radial quantum number
and $n_c$ is the radial quantum number cutoff, such that we include only
the single particle states with $\omega(2n+\ell+3/2) \le \epsilon_c = \omega(n_c + 3/2)$.
In Eq.~\ref{eqn:bdg.matrix}, the diagonal matrix element is
$
K_{\sigma, \ell}^{n,n'} = \xi_{n,\ell}^\sigma \delta_{n,n'}
$
where $\delta_{i,j}$ is the Kronecker delta, and the off-diagonal matrix element is
$
\Delta_\ell^{n,n'} \approx \int r^2 dr \Delta(r) R_{n,\ell}(r) R_{n',\ell}(r).
$
Although $\Delta(\mathbf{r})$ becomes axially symmetric when $\Omega \ne 0$,
it is convenient to define
$
\Delta(r) = \int d\Omega_\mathbf{r} \Delta(\mathbf{r})/(4\pi)
$
leading to
\begin{eqnarray}
\label{eqn:op}
\Delta(r) = -\frac{g}{4\pi} \sum_{\ell,m,\gamma,n,n'} 
\widetilde{R}^\uparrow_{\ell,m,\gamma,n}(r) \widetilde{R}^\downarrow_{\ell,m,\gamma,n'}(r) f(\epsilon_{\ell,m,\gamma}),
\end{eqnarray}
where we introduced
$
\widetilde{R}^\uparrow_{\ell,m,\gamma,n}(r) = c_{\ell,m,\gamma,n} R_{n,\ell}(r)
$
and
$
\widetilde{R}^\downarrow_{\ell,m,\gamma,n}(r) = d_{\ell,m,\gamma,n} R_{n,\ell}(r).
$
Similarly, we define
$n_\sigma(r) = \int d\Omega_\mathbf{r} n_\sigma(\mathbf{r})/(4\pi)$ leading to
\begin{eqnarray}
\label{eqn:n}
n_\sigma(r) = \frac{1}{4\pi}\sum_{\ell,m,\gamma,n,n'} 
\widetilde{R}^\sigma_{\ell,m,\gamma,n}(r) \widetilde{R}^\sigma_{\ell,m,\gamma,n'}(r) f(s_\sigma\epsilon_{\ell,m,\gamma}).
\end{eqnarray}
Lastly, $N_\sigma$ reduces to
$
N_\uparrow = \sum_{\ell,m,\gamma,n} c_{\ell,m,\gamma,n}^2 f(\epsilon_{\ell,m,\gamma}) 
$
and
$
N_\downarrow = \sum_{\ell,m,\gamma,n} d_{\ell,m,\gamma,n}^2 f(-\epsilon_{\ell,m,\gamma}).
$
When $\Omega \to 0$,  the eigenfunction coefficients $c_{\ell,m,\gamma,n}$ 
and $d_{\ell,m,\gamma,n}$ and the eigenvalues $\epsilon_{\ell,m,\gamma}$ 
become independent of $m$, and Eqs.~(\ref{eqn:bdg.matrix}),~(\ref{eqn:op}) 
and~(\ref{eqn:n}) reduce to the usual ones, see \textit{e.g.} 
Ref.~\cite{ohashi, castorina, mizushima, torma}. Therefore, due to the coupling between 
different $m$ states, the rotating case is numerically much more involved 
compared to the nonrotating case.

In addition to $n_\sigma(r)$, we want to calculate the local density of normal (rotating) 
fermions $n_{\sigma,N}(r)$. For this purpose, we use the local current density 
$
\mathbf{J_\sigma} (\mathbf{r})  = \langle \mathbf{{\cal J}_\sigma} (\mathbf{r}) \rangle,
$ 
where
$
\mathbf{{\cal J}_\sigma} (\mathbf{r}) = [1/(2M i)]
[\psi_\sigma^\dagger(\mathbf{r}) \nabla \psi_\sigma(\mathbf{r}) - H.c.]
$
is the quantum mechanical probability current operator and $H.c.$ is the Hermitian conjugate.
This leads to
$
\mathbf{J_\uparrow} (\mathbf{r}) = [1/(2M i)] \sum_{\eta}
[u_\eta^*(\mathbf{r}) \nabla u_\eta(\mathbf{r}) f(\epsilon_\eta) - H.c.]
$
and
$
\mathbf{J_\downarrow} (\mathbf{r}) = [1/(2M i)] \sum_{\eta}
[v_\eta(\mathbf{r}) \nabla v_\eta^*(\mathbf{r}) f(-\epsilon_\eta) - H.c.].
$
The current, similar to the classical case, can be written as 
$
\mathbf{J_\sigma} (\mathbf{r}) = n_{\sigma,N}(\mathbf{r}) \mathbf{v} (\mathbf{r}),
$
where $n_{\sigma,N}(\mathbf{r})$ is the local density and 
$
\mathbf{v} (\mathbf{r}) = \Omega \mathbf{\widehat{z}} \times \mathbf{r}
$ 
is the local velocity of normal fermions corresponding to a rigid-body rotation. 
Since the normal fermions are expelled towards the trap edge, we approximate
$
n_{\sigma,N}(r) = \int d\Omega_\mathbf{r} n_{\sigma,N}(\mathbf{r})/(4\pi)
$
as
\begin{align}
\label{eqn:cd}
n_{\sigma,N}(r) \approx \frac{s_\sigma}{4\pi M \Omega r^2} & \sum_{\ell,m,\gamma,n,n'} m
\widetilde{R}^\sigma_{\ell,m,\gamma,n}(r) \nonumber \\
& \widetilde{R}^\sigma_{\ell,m,\gamma,n'}(r) f(s_\sigma\epsilon_{\ell,m,\gamma}),
\end{align}
such that $J_\sigma(r) = \int d\Omega_\mathbf{r} J_{\sigma}(\mathbf{r})/(4\pi) \sim \Omega r n_{\sigma,N}(r)$.

Having discussed the BdG formalism, next, we analyze the ground state ($T = 0$) phases
for both population balanced ($N_\uparrow = N_\downarrow$ or $\mu_\uparrow = \mu_\downarrow$) 
and imbalanced ($N_\uparrow \ne N_\downarrow$ or $\mu_\uparrow \ne \mu_\downarrow$) Fermi gases.
This is achieved by solving the BdG equations~(\ref{eqn:bdg.matrix}),~(\ref{eqn:op}) 
and~(\ref{eqn:n}) self-consistently as a function of the dimensionless 
parameter $1/(k_F a_F)$ where $k_F$ is the Fermi momentum defined via the Fermi energy 
\begin{equation}
\epsilon_F = \omega (n_F + 3/2)  = \frac{k_F^2}{2M} = \frac{1}{2}M \omega^2 r_F^2 \approx \omega (3N)^{1/3}
\end{equation}
and $N = N_\uparrow + N_\downarrow = (n_F + 1)(n_F + 2)(n_F + 3)/3$. Here, $n_F$ 
and $r_F$ are the corresponding Fermi level and Thomas-Fermi radius, respectively.
In our numerical calculations, we choose $n_F = 15$ and $n_c = 65$, which corresponds 
to a total of $N = 1632$ fermions and $\epsilon_c \approx 4\epsilon_F$, respectively. 
Here, it is important to emphasize that we do not expect any qualitative change in our 
results with higher values of $n_F$ and/or $n_c$, except for minor quantitative variations.

\begin{figure} [htb]
\centerline{\scalebox{0.45}{\includegraphics{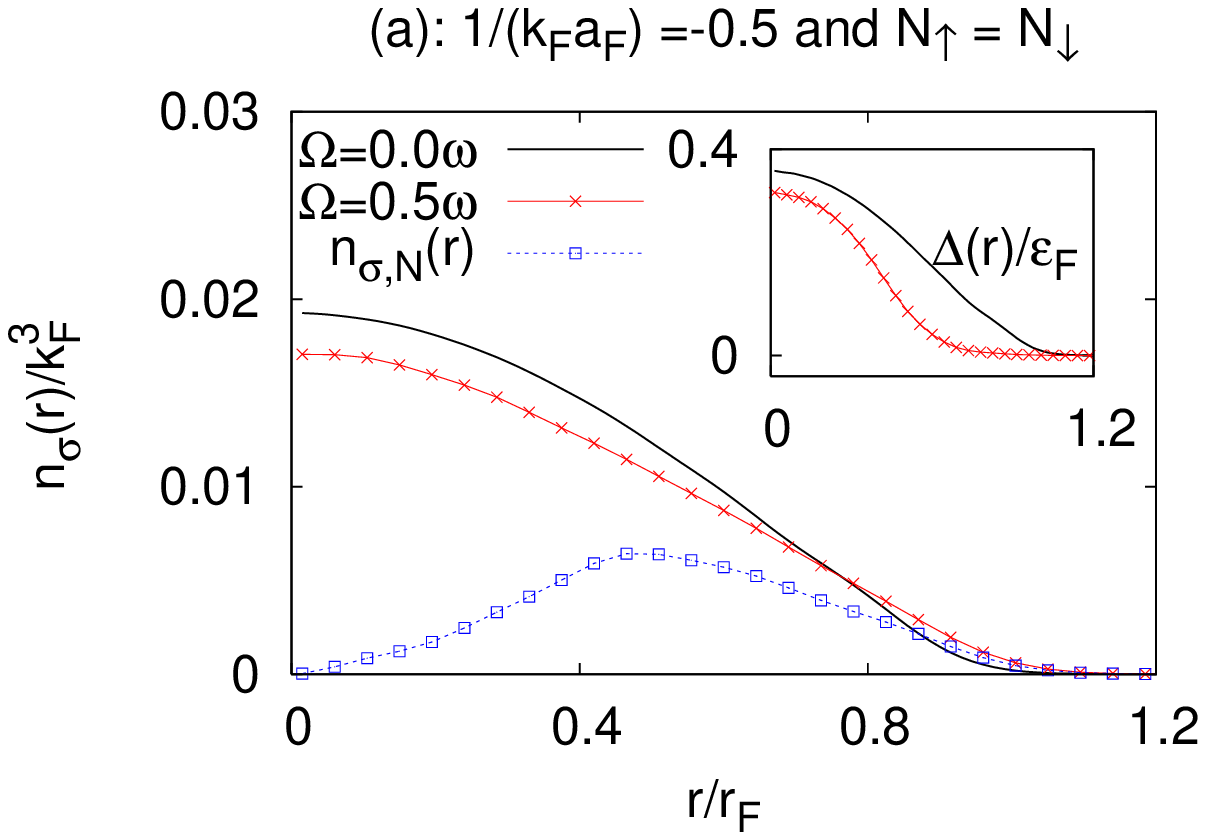}}}
\centerline{\scalebox{0.45}{\includegraphics{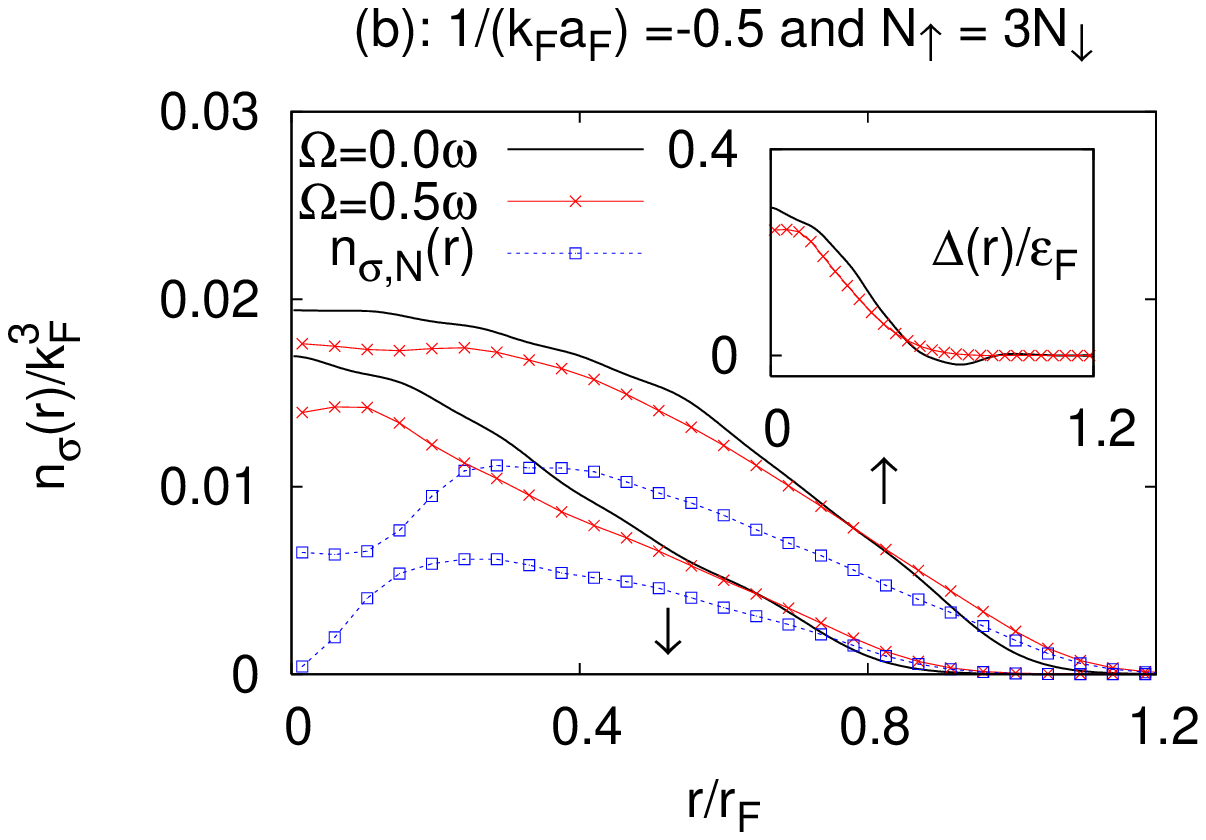}}}
\caption{\label{fig:density-BCS}
We compare the density $n_\sigma(r)$ and the superfluid order parameter 
$\Delta(r)$ for nonrotating ($\Omega = 0$) and rotating 
($\Omega = 0.5 \omega$) Fermi gases. 
Here, $1/(k_F a_F) = -0.5$, and $N_\uparrow = N_\downarrow$ in (a) 
and $N_\uparrow = 3N_\downarrow$ in (b).
We also show the density $n_{\sigma,N}(r)$ of normal fermions 
(squared-dotted line) for the rotating case.
}
\end{figure}

In Fig.~\ref{fig:density-BCS}, we consider a weakly interacting Fermi gas on the BCS 
side with $1/(k_F a_F) = -0.5$, and show $n_\sigma(r)$ and $\Delta(r)$ 
for nonrotating ($\Omega = 0$) and rotating ($\Omega = 0.5 \omega$) cases. 
When $N_\uparrow = N_\downarrow$, we find that $\Delta(r)$ depletes 
everywhere inside the trap and especially around the trap edge.
This is because it is easier to break the Cooper pairs that are located
towards the trap edge in comparison to the ones that occupy the center (see below).
Our result is quantitatively different from the LDA one where $\Delta(r)$
depletes only around the trap edge~\cite{urban}.
When $N_\uparrow = 3N_\downarrow$, in addition to such an effect, 
the spatial modulation of $\Delta(r)$ disappears in the rotating case as 
shown in Fig.~\ref{fig:density-BCS}(b). The depletion of $\Delta(r)$ leads to 
a decrease (increase) in $n_\sigma(r)$ around the trap center (edge) due 
to the centrifugal force caused by the rotation. It also leads to a phase separation 
between the nonrotating fully paired superfluid (FPS) atoms located around 
the trap center and the rigidly rotating normal (nonpaired) ones located 
towards the trap edge, with a coexistence region in between, which is in 
agreement with the LDA result~\cite{urban}. We characterize 
the FPS, coexistence and normal phases by $n_\sigma(r) \gg n_{\sigma, N}(r) = 0$,
$n_\sigma(r) \gtrsim n_{\sigma, N}(r) \ne 0$ and $n_\sigma(r) = n_{\sigma, N}(r) \ne 0$, 
respectively. Notice that $\Delta(r)$ decreases smoothly as a function of $r$ from the 
FPS to the normal region where it vanishes, which is in contrast with the
LDA result where $\Delta(r)$ is nonanalytic at the transition~\cite{urban}.
Since $\Delta(r)$ is finite in the coexistence region, this region corresponds to a 
partially paired superfluid (PPS), and it occupies a larger region compared
to the LDA resuls~\cite{urban}. In addition, the trap center becomes a PPS 
for $\Omega \gtrsim 0.5 \omega$ when $1/(k_Fa_F) = -0.5$.

\begin{figure} [htb]
\centerline{\scalebox{0.45}{\includegraphics{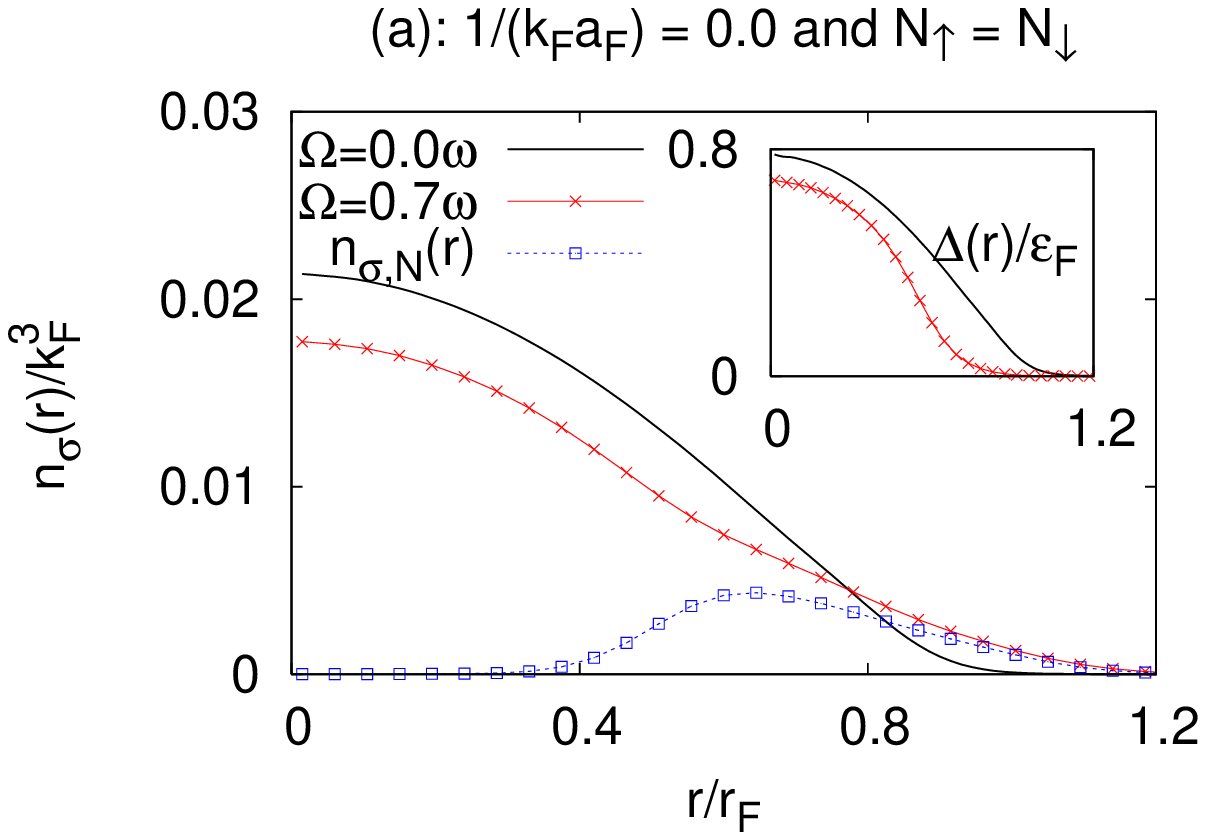}}}
\centerline{\scalebox{0.45}{\includegraphics{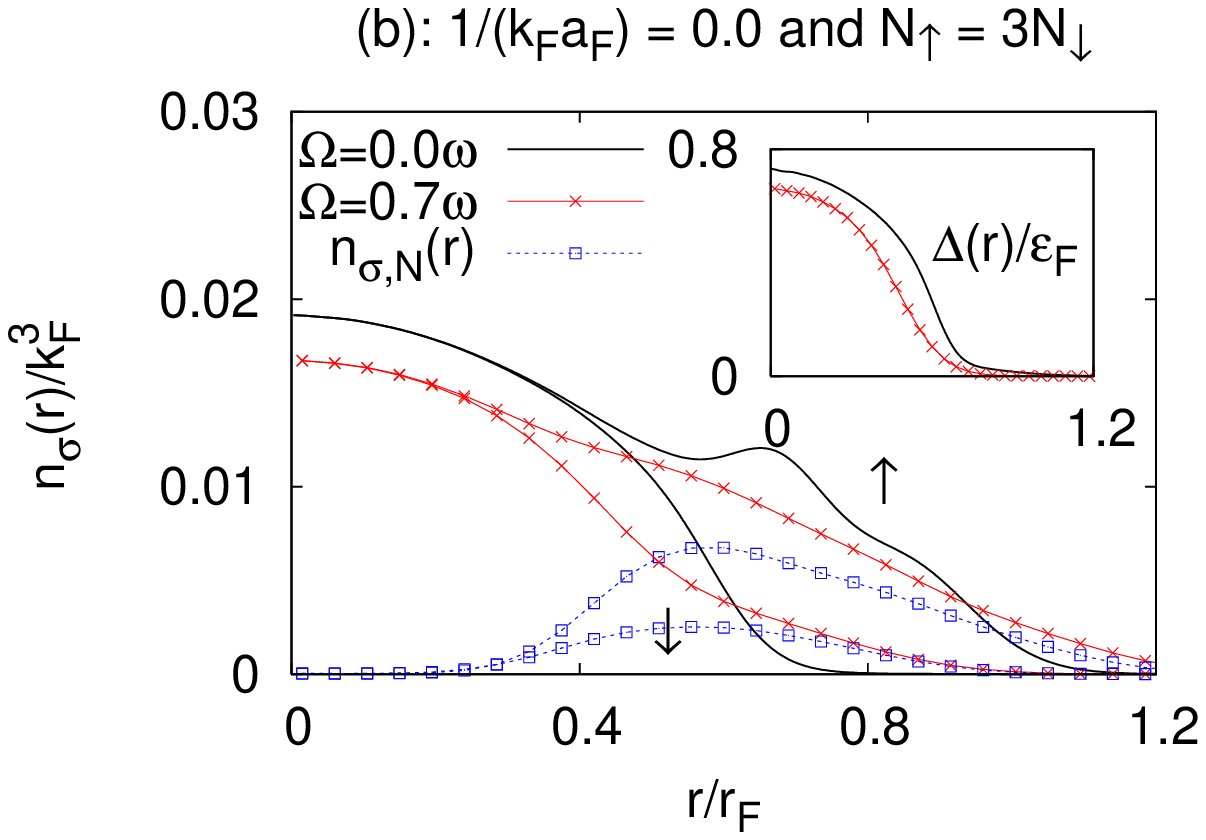}}}
\caption{\label{fig:density-resonance}
We compare the density $n_\sigma(r)$ and the superfluid order parameter 
$\Delta(r)$ for nonrotating ($\Omega = 0$) and rotating 
($\Omega = 0.7 \omega$) Fermi gases. 
Here, $1/(k_F a_F) = 0.0$, and $N_\uparrow = N_\downarrow$ in (a) 
and $N_\uparrow = 3N_\downarrow$ in (b).
We also show the density $n_{\sigma,N}(r)$ of normal fermions 
(squared-dotted line) for the rotating case.
}
\end{figure}

In Fig.~\ref{fig:density-resonance}, we consider a strongly interacting Fermi gas
at unitarity with $1/(k_F a_F) = 0$, and show $n_\sigma(r)$ and $\Delta(r)$ 
for nonrotating ($\Omega = 0$) and rotating ($\Omega = 0.7 \omega$) cases. 
The main effects of rotation are
qualitatively similar to the weakly interacting case. However, since the Cooper 
pairs become more strongly bound as a function of the interaction strength, 
it requires much faster $\Omega$ to break them. For instance, at unitarity, 
the entire superfluid is robust for $\Omega \lesssim 0.3\omega$, and the trap 
center stays as an FPS even for $\Omega \sim \omega$ (not shown). 
Therefore, the effects of rotation become weaker as the interaction strength
increases, and both the PPS and the normal regions eventually disappear 
in the molecular limit (not shown), \textit{i.e.} rotation can not break 
any Cooper pair in the molecular limit.

Another important observable is the local angular momentum defined by
$
L_{z,\sigma} (\mathbf{r}) =  \langle \psi_\sigma^\dagger(\mathbf{r}) {\cal L}_z \psi_\sigma(\mathbf{r}) \rangle.
$
This leads to
$
L_{z,\uparrow} (\mathbf{r}) = -i\sum_{\eta} u_\eta^*(\mathbf{r}) \partial u_\eta(\mathbf{r})/\partial \varphi_{\mathbf{r}} f(\epsilon_\eta)
$
and
$
L_{z,\downarrow} (\mathbf{r}) = -i\sum_{\eta} v_\eta(\mathbf{r}) \partial v_\eta^*(\mathbf{r})/\partial \varphi_{\mathbf{r}} f(-\epsilon_\eta).
$
Since the superfluid atoms do not carry angular momentum, $L_{z,\sigma} (\mathbf{r})$ 
is directly related to the local density of normal fermions via 
$
L_{z,\sigma}(r) = \int d\Omega_\mathbf{r} L_{z,\sigma}(\mathbf{r})/(4\pi) \approx M \Omega r^2 n_{\sigma, N} (r).
$
Therefore, $L_{z,\sigma}(r) $ can be easily extracted from Figs.~\ref{fig:density-BCS} 
and~\ref{fig:density-resonance}. For completeness, the total angular momentum 
$L_{z,\sigma} = \int d\mathbf{r} L_{z,\sigma}(\mathbf{r})$ becomes
$
L_{z,\uparrow} = \sum_{\ell,m,\gamma,n,n'} m c_{\ell,m,\gamma,n}^2 f(\epsilon_{\ell,m,\gamma})
$
and
$
L_{z,\downarrow} = -\sum_{\ell,m,\gamma,n,n'} m d_{\ell,m,\gamma,n}^2 f(-\epsilon_{\ell,m,\gamma}),
$
and it increases with increasing $\Omega$. In the weakly attracting limit, 
$L_{z,\sigma}$ becomes its rigid-body value when $\Omega$ increases
high enough so that $\Delta(r) \to 0$ everywhere (not shown).

\begin{figure} [htb]
\centerline{\scalebox{0.3}{\includegraphics{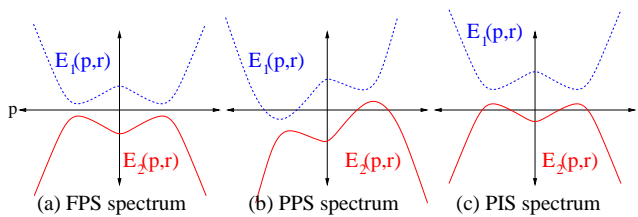}}}
\caption{\label{fig:topology}
Schematic diagrams showing the excitation spectrum of 
(a) a gapped fully paired superfluid (FPS) phase at $r = 0$,
(b) a gapless partially paired superfluid (PPS) phase at $r \ge r_T$,
and (c) a gapless population imbalanced superfluid (PIS) phase at $r = 0$
as a function of momentum $\mathbf{p}$.
}
\end{figure}

The microscopic mechanism responsible for the pair breaking effects can be
understood analytically within the semi-classical LDA, \textit{i.e.} each component
of the Fermi gas is considered as locally homogenous at each position 
$\mathbf{r}$ with a local chemical potential $\mu_\sigma(\mathbf{r})$. In this approximation, 
the local quasiparticle and quasihole excitation branches are~\cite{urban, iskin-melo}
\begin{eqnarray}
E_{1,2}(\mathbf{p}, \mathbf{r}) = \frac{\mu_\uparrow - \mu_\downarrow}{2} + \mathbf{v}(\mathbf{r}) \cdot \mathbf{p} \pm E_0(p,\mathbf{r}),
\end{eqnarray}
where $E_0(p,\mathbf{r}) = \sqrt{[p^2/(2M) - \mu(r)]^2 + \Delta^2(\mathbf{r})}$ is the usual spectrum for
nonrotating and population balanced mixtures,
$\mathbf{v}(\mathbf{r}) = \Omega \widehat{\mathbf{z}} \times \mathbf{r}$ is the
velocity and $\mathbf{p}$ is the momentum. 
In Fig.~\ref{fig:topology}, we present
three schematic diagrams showing $E_1(\mathbf{p}, \mathbf{r})$ and 
$E_2(\mathbf{p}, \mathbf{r}) $ as a function of $\mathbf{p}$ for fixed values of 
$\mathbf{r}$. At each $\mathbf{r}$, the many-body ground state wavefunction 
fills up all of the states with negative energy, and excitations correspond to removing 
a quasiparticle or a quasihole from a filled state and adding it to the one that is not filled.  
In order to show that these excitation spectra correspond to three topologically 
distinct superfluid phases that can be observed in atomic systems, next we discuss
$E_{1,2}(\mathbf{p}, \mathbf{r})$ in the $z = 0$ plane such that $\mathbf{r} \equiv (x,y,0)$.

First, we consider the population balanced case where $\mu_\uparrow = \mu_\downarrow$.
In the FPS phase, the excitation spectrum is symmetric around the zero energy 
axis, \textit{i.e.} $E_{1}(\mathbf{p}, \mathbf{r}) = -E_2(\mathbf{p}, \mathbf{r})$,
leading to a gapped spectrum as shown in Fig.~\ref{fig:topology}(a). 
However, in the rotating case, there is a local asymmetry between the pairing states 
$\{ \mathbf{r}, \mathbf{p}, \uparrow; \mathbf{r}, -\mathbf{p}, \downarrow \}$ and 
$\{ \mathbf{r}, -\mathbf{p}, \uparrow; \mathbf{r}, \mathbf{p}, \downarrow \}$. 
When this asymmetry becomes sufficiently large, there exists a momentum 
space region $p_- \le p \le p_+$ where $E_1(-\mathbf{p}, \mathbf{r}) \le 0$ 
and $E_2(\mathbf{p}, \mathbf{r}) \ge 0$. 
This occurs for position space region $r \ge r_T$ when the condition
$
A(r) = 2M \Omega^2 r^2 [\mu(r) + M \Omega^2 r^2/2] - \Delta^2(r) \ge 0
$
is satisfied, where $r_T$ is defined through $A(r_T) = 0$ and
$
p_\pm = 2M [\mu(r) + M \Omega^2 r^2]  \pm 2M\sqrt{A(r)}.
$
Therefore, both $E_{1,2}(\mathbf{p}, \mathbf{r})$ have two zeros at $p = p_\pm$, 
and the excitation spectrum becomes gapless at these momenta. 
We characterize this phase as the PPS, and its excitation spectrum 
is shown in Fig.~\ref{fig:topology}(b). 

For a weakly attracting gas, the condition $A(r) \ge 0$ can only be 
satisfied near the trap edge when $\Omega \ll \omega$, but it can also be satisfied 
near the trap center when $\Omega \sim \omega$. However, in the strongly 
interacting limit where $\mu$ is small yet positive, this condition 
can only be satisfied for sufficiently small values of $\Delta(r)$ near the trap 
edge even when $\Omega \sim \omega$. Finally, in the molecular limit 
where $\mu$ is negative, this condition can not be satisfied 
anywhere inside the trap, leading to a superfluid phase with a gapped excitation 
spectrum, \textit{i.e.} the Cooper pairs are robust in the molecular limit.
As one may expect, the asymmetric pairing caused by the rotation does not 
lead to a local population imbalance at any position $r$ 
since $E_1(-\mathbf{p}, \mathbf{r}) \le 0$ when 
$E_2(\mathbf{p}, \mathbf{r}) \ge 0$ and vice versa. However, this asymmetry
prevents the formation of Cooper pairs in the phase space region when
both $E_{1,2}(\mathbf{p}, \mathbf{r}) \ge 0$ or both $E_{1,2}(\mathbf{p}, \mathbf{r}) \le 0$, 
and thus it is responsible for the creation of the PPS and the normal phases.

We find that the local excitation spectrum changes from gapped (FPS) 
to gapless (PPS) at position $r = r_T$. In homogenous (infinite) systems 
such a change is classified as a topological quantum phase transition~\cite{volovik}.
Since this change occurs in the momentum space, 
it does not leave any strong signature in the momentum averaged 
observables such as $\Delta(r)$, $n(r)$, etc. However, its direct consequences 
can be observed via a recently developed position and momentum 
resolved spectrocopy~\cite{stewart}. For instance, the local 
density distributions~\cite{iskin-melo}
\begin{eqnarray*}
n_{\uparrow,\downarrow} (\mathbf{p},\mathbf{r}) = \widetilde{u}^2(\mathbf{p}, \mathbf{r}) f[\pm E_{1,2}(\mathbf{p}, \mathbf{r})]
+ \widetilde{v}^2(\mathbf{p}, \mathbf{r}) f[\pm E_{2,1}(\mathbf{p}, \mathbf{r})], 
\end{eqnarray*}
are nonanalytic at $p = p_\pm$ when $r \ge r_T$. Here, 
$
\widetilde{u}^2(\mathbf{p}, \mathbf{r}) = \{1 + [\epsilon(p) - \mu(r)]/E_0(\mathbf{p}, \mathbf{r})] \}/2
$
and
$
\widetilde{v}^2(\mathbf{p}, \mathbf{r}) = \{1 - [\epsilon(p) - \mu(r)]/E_0(\mathbf{p}, \mathbf{r})] \}/2
$
are the usual coherence factors with $\epsilon(p) = p^2/(2M)$.

For population imbalanced superfluids (PIS) where $\mu_\uparrow \ne \mu_\downarrow$, 
the excitation branches shift upwards (downwards), and one of them cross 
zero energy axis when $N_\uparrow > N_\downarrow$ ($N_\uparrow < N_\downarrow$),
leading to a gapless excitation spectrum.
This is expected since the excess fermions can only exist in regions with 
both $E_{1,2}(\mathbf{p}, \mathbf{r}) \ge 0$ or both $E_{1,2}(\mathbf{p}, \mathbf{r}) \le 0$.
In the absence of rotation~\cite{iskin-melo}, for a weakly attracting PIS 
only one of the excitation branches has four zeros as shown 
in Fig.~\ref{fig:topology}(c) for the $N_\uparrow > N_\downarrow$ case.
However, in the strongly attracting limit, this branch has only two zeros (not shown).
When the system is rotating, the excitation spectra tilt 
similar to that shown in Fig.~\ref{fig:topology}(b) (not shown). 
We remark in passing that a similar quantum phase transition 
with its experimental signatures has recently been discussed in the context of 
trapped $p$-wave superfluids~\cite{iskin-williams}.

To conclude, we used the BdG formalism to analyze the effects of adiabatic 
rotation on the ground state phases of harmonically trapped Fermi gases.
We found that the rotation breaks Cooper pairs that are located near the trap edge, 
and that this leads to a phase separation between the nonrotating superfluid 
(fully paired) atoms located around the trap center and the rigidly rotating normal 
(nonpaired) atoms located towards the trap edge with a coexistence 
(partially paired) region in between. We also showed that the rotation 
reveals a topological quantum phase transition in the momentum space 
as a function of radial distance. An interesting extention of our work
is to study emergence of dynamic instabilities for fast enough 
rotation~\cite{stringari, tonini}.

\end{document}